\documentclass[conference,final]{IEEEtran}

\usepackage[T1]{fontenc}
\usepackage[USenglish]{babel}	%
\usepackage[numbers]{natbib}
	
\usepackage[final]{graphicx}
	\graphicspath{{figures/}}
	\DeclareGraphicsExtensions{.pdf,.eps,.png,.jpg}
\usepackage[caption=false]{subfig}
\usepackage{amsmath,amssymb}
\usepackage{bm}	%
\usepackage[binary-units,per-mode=reciprocal]{siunitx}
\usepackage{booktabs}
\usepackage{collcell}
\usepackage{multicol}
\usepackage{spreadtab}
\usepackage{paralist}
\usepackage{enumitem}
	\newlist{enummethod}{enumerate}{1}
	\setlist[enummethod,1]{label=\arabic*.,
                   ref  =\arabic*}
\usepackage{ifdraft}
\usepackage{pbox}
\usepackage[obeyFinal]{todonotes}
\usepackage[nolist,smaller]{eee-acronyms}
	\let\Acs\acs
\usepackage{acronym-patch,acronym-abstract}
\usepackage{acronym-alwayslong}
	\acronymalwayslong{AI}
	\acronymalwayslong{HDC}
	\acronymalwayslong{LA}
	\acronymalwayslong{MEP}
	\acronymalwayslong{MPPT}
	\acronymalwayslong{PVT}
	\acronymalwayslong{STA}
	\acronymalwaysshort{CAD}
	\acronymalwaysshort{RTL}
\usepackage{eee-macros}
\csname @namedef\endcsname{processlibname@umc65ll}{\pnoun{UMC \SI{65}{\nano\metre} Low-leakage}}
\def\processlibname#1{%
	\expandafter\ifx\csname processlibname@#1\endcsname\relax
		\expandafter\errmessage{Library name `#1' didn't match any that I know.}\errorstopmode
	\else
		\csname processlibname@#1\endcsname%
	\fi
}
\csname @namedef\endcsname{processlibnameshort@umc65ll}{\pnoun{UMC\num{65}LL}}
\def\processlibnameshort#1{%
	\expandafter\ifx\csname processlibnameshort@#1\endcsname\relax
		\expandafter\errmessage{Library name `#1' didn't match any that I know.}\errorstopmode
	\else
		\csname processlibnameshort@#1\endcsname%
	\fi
}

\usepackage{boolean}
\usepackage{makecell}

\providecommand{\tikzdep}[1]{\relax}%
\providecommand{\tikzinput}[1]{\includegraphics{tikzfigure-#1}}%

\newcommand*{\feat}[1]{f}
\newcommand*{\finput}{feature input}
\newcommand*{\Finput}{Feature input}
\newcommand*{\finputs}{\finput{}s}
\newcommand*{\Finputs}{\Finput{}s}
\newcommand*{\Ninputs}{\ensuremath{N_\textrm{Inputs}}}
\newcommand*{\Nclauses}{\ensuremath{N_\textrm{Clauses}}}
\newcommand*{\Nclasses}{\ensuremath{N_\textrm{Classes}}}
\newcommand*{\Ntas}{\ensuremath{N_\textrm{TAs}}}

\newcommand*{\fbstage}[1]{\text{\textsmaller{FB#1}}}
\newcommand*{\fbrail}[2]{\xrail{\fbstage{#1}}{\fbtype{#2}}}
\newcommand*{\fbraila}[2]{\xrail{\fbstage{#1}}{\tasaction{#2}}}
\newcommand*{\fbtype}[1]{%
  \def\tempa{#1}%
  \def\tempb{0}%
  \text{%
  \ifx\tempa\tempb
    none%
  \else
    \textsmaller{T#1}%
  \fi
  }%
}
\newcommand*{\fbprobsig}[1]{%
  \def\tempa{#1}%
  \def\tempb{2}%
  \def\tempc{3}%
  \ifx\tempa\tempb
    \signal[\mathsmaller{\tparam{}}]{p}%
  \else
    \ifx\tempa\tempc
      \signal[\sparam{}]{p}%
    \else
      \signal[#1]{p}%
    \fi
  \fi
}
\newcommand*{\prob}[1]{\ensuremath{%
    P{\textstyle{(#1)}}%
}}
\newcommand*{\fbprob}[1]{\prob{\fbtype{#1}}}
\newcommand*{\taaction}[1]{%
  \def\tempa{#1}%
  \def\tempb{0}%
  \def\tempc{1}%
  \text{%
  \ifx\tempa\tempb
    inaction%
  \else
    \ifx\tempa\tempc
      penalty%
    \else
      reward%
    \fi
  \fi
  }%
}
\newcommand*{\tasaction}[1]{%
  \def\tempa{#1}%
  \def\tempb{0}%
  \def\tempc{1}%
  \text{%
  \ifx\tempa\tempb
      i%
  \else
    \ifx\tempa\tempc
      p%
    \else
      r%
    \fi
  \fi
  }%
}

\newcommand*{\csum}{\signal[sum]{c}}
\newcommand*{\cneg}{\ensuremath{\text{c}_\text{neg}}}
\newcommand*{\tparam}{\ensuremath{T}}
\newcommand*{\sparam}{\ensuremath{s}}

\newcommand*{\yexp}{\signal{yexp}}

\newcommand*{\transplus}[1]{\signal[^\mathsmaller{+}]{#1}}
\newcommand*{\transminus}[1]{\signal[^\mathsmaller{-}]{#1}}

\newcommand{\pnoun}[1]{\textsc{#1}}
\newcommand*{\signal}[2][]{\ensuremath{\mathsf{#2}_\mathsf{#1}}}
\newcommand*{\nsignal}[2][]{\ensuremath{\bnot{\mathsf{#2}_\mathsf{#1}}}}
\newcommand*{\prail}[2][]{\ensuremath{\signal[#1]{#2}^\mathsf{p}}}
\newcommand*{\nrail}[2][]{\ensuremath{\signal[#1]{#2}^\mathsf{n}}}
\newcommand*{\xrail}[3][]{\ensuremath{\signal[#1]{#2}^\mathsf{#3}}}
\let\oldvec\vec
\renewcommand*{\vec}[2][]{\ensuremath{\oldvec{\mathsf{#2}}_{#1}}}

\newcommand{\cels}{C-elements}

\newcommand{\dr}{dual-rail}

\newcommand{\popcount}{\popcounta}
\newcommand{\Popcount}{\Popcounta}
\newcommand{\PopCount}{\PopCounta}

\newcommand{\popcounta}{population count}
\newcommand{\Popcounta}{Population count}
\newcommand{\PopCounta}{Population Count}

\renewcommand{\vdd}[1]{%
        \def\temp{#1}\ifx\temp\empty
                \ensuremath{V_{\mathrm{\textsc{dd}}}}%
        \else
                \ensuremath{V_{\mathrm{\textsc{dd},#1}}}%
        \fi
}

\newcommand{\breakingspace}[1]{#1\hspace{0pt}}
\newcommand{\trans}{\texorpdfstring{\ignorespaces\,$\to$\breakingspace{\,}\ignorespaces}{ to }}
\newcommand{\sptocw}{spacer\trans{}codeword}

\let\oldgate\gate
\renewcommand{\gate}[1]{\oldgate{#1} gate}
\usepackage{suffix}
\WithSuffix\newcommand\gate*{\oldgate}
\newcommand*\gates[1]{\gate{#1}s}
\newcommand*\nth[2][th]{$#2\text{#1}$}
\newcommand*{\bname}[2][]{$\textsc{#2}_{#1}$}     %

\newcolumntype{L}{>{$}l<{$}}%
\newcolumntype{R}{>{$}r<{$}}%
\newcolumntype{C}{>{$}c<{$}}%

\usepackage[noabbrev,capitalise]{cleveref} %
	\crefname{enummethodi}{Method}{Methods}

\newcommand*\stgscale{0.235}
\newcommand*\petriscale{0.3}

\begin{document}
	\bstctlcite{IEEE:BSTcontrol}
	\title{Self-timed Reinforcement Learning\\using Tsetlin Machine}
	\author{%
		\IEEEauthorblockN{Adrian~Wheeldon, Alex~Yakovlev and Rishad~Shafik}
		\IEEEauthorblockA{
			Microsystems Group, Newcastle University, UK\\
			Email: \{a.r.wheeldon2, alex.yakovlev, rishad.shafik\}@ncl.ac.uk
		}
	}
	\maketitle
	\ifoptionfinal{}{\thispagestyle{plain}\pagestyle{plain}}%

	\begin{abstract}

  We present a hardware design for the learning datapath of the \ac{TM}
  algorithm, along with a latency analysis of the inference datapath.
  In order to generate a low energy hardware which is suitable for pervasive
  \ac{AI} applications, we use a mixture of asynchronous design
  techniques---including Petri nets, \acp{STG}, \ac{DR} and \ac{BD}.
  The work builds on previous design of the inference hardware, and includes an
  in-depth breakdown of the automaton feedback, probability generation and
  \acp{TA}.
  Results illustrate the advantages of asynchronous design in applications such
  as personalized healthcare and battery-powered \ac{IoT} devices, where energy is
  limited and latency is an important figure of merit.
  Challenges of \ac{STA} in asynchronous circuits are also addressed.

\end{abstract}

	\section{Introduction}\label{sec:intro}

We present a comprehensive design and analysis for an asynchronous
\emph{learning} datapath based
on the \ac{TM} algorithm. When coupled with asynchronous inference
hardware~\cite{wheeldon2021lowlatency}, a complete asynchronous \ac{TM} capable
of online learning is formed. We also analyze the latency of the inference
datapath, showing its evolution during training.

The \ac{TM}~\cite{granmo2018tsetlin} algorithm is effective in many large
classification
problems~\cite{berge2018using,blakely2020closedform,bhattarai2020measuring}. In
addition, the \ac{TM}'s reinforcement learning and logic-based inference make it
a good candidate for energy efficient hardware~\cite{wheeldon2020learning}. We
design an energy-frugal \ac{TM} hardware with a view of use cases in pervasive
\ac{AI}; \eg{} in personal healthcare, accessibility, environmental monitoring
and predictive maintenance.
We give a brief introduction to the \ac{TM} algorithm in \cref{sec:tm-intro}.

We wish to implement the \ac{TM} using asynchronous circuits as they can be
beneficial for low energy sensing systems~\cite{wheeldon2019selftimed}, when
tightly coupled with analog blocks~\cite{sokolov2020automating}, and when
power delivery is unstable or unpredictable~\cite{zhang2011novel}. These
features are often present in the aforementioned
applications~\cite{balsamo2017wearable}.

The architecture of the hardware is designed using a hybrid Petri net model
in \cref{sec:tm}. The model incorporates inference and learning components.

\Acp{TM} use groups of reinforcement automata, called \acp{TA}, to create an ensemble
learning effect. We decompose the automata reinforcement feedback into three
stages to aid the hardware design in \cref{sec:fb}.

The reinforcement process of the \ac{TM} involves random choice with defined
probabilities---some fixed and some varying at runtime. These probabilistic choices
enable diversity of learning in the \acp{TA}. We design \iac{QDI} \ac{PRBG} for
this use in the \ac{TM} in \cref{sec:rand}, and motivate its use
in other low power applications.

Considering the implementation of the state-holding \acp{TA}, we compare \ac{QDI}
and \ac{BD} design styles suitable for the low energy applications in
\cref{sec:ta}. The comparison will show the limitations of the two asynchronous
design methods in terms of area, energy and scalability.
We use the \pnoun{Workcraft} framework~\cite{workcraft} for specification and
synthesis of the \ac{QDI} implementation using \acp{STG}, as well as
verification of the \ac{STG} properties required for a functional and robust
design.
For \ac{BD}, we are inspired by the desynchronization workflow of
\citet{cortadella2006desynchronization} in using a commercial synthesis tool
designed for synchronous design. \Ac{BD} circuits are
readily-implementable with industrial-quality tools, in contrast with \ac{QDI}
circuits.

\emph{Main contributions of this work:}
\begin{itemize}
  \item Visualization of an asynchronous \ac{TM} hardware architecture using Petri nets (\cref{sec:tm}).
  \item Latency analysis of the asynchronous \ac{TM} inference datapath from
    \cite{wheeldon2021lowlatency} and its evolution during training (\cref{sec:tm}).
  \item A detailed decomposition of \ac{TA} reinforcement feedback into three stages (\cref{sec:fb}).
  \item Design of \iac{QDI} \ac{PRBG} for probabilistic choice in the \ac{TM} and other low power applications (\cref{sec:rand}).
  \item Comparison of \ac{QDI} and \ac{BD} implementations of the \ac{TA} for low energy hardware (\cref{sec:ta}).
\end{itemize}

	\section{Tsetlin Machines in Brief}\label{sec:tm-intro}

\begin{figure}
        \centering
        \tikzinput{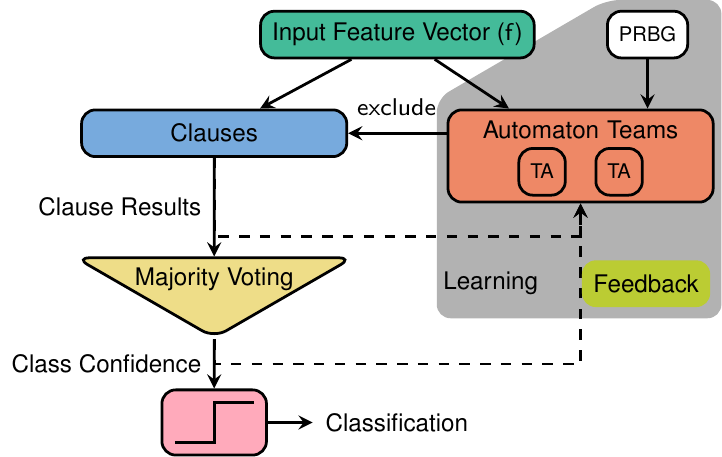}
        \caption{%
                Overview of the \acf{TM} architecture.
        }\label{fig:tm-overview}
\end{figure}

\Acp{TM} learn patterns in binary data using
propositional logic. The main inference component of the \ac{TM} is the
\emph{clause} that composes an \gate*{and} expression of the input features and
their complements. The \ac{TM} comprises many clauses, each producing a vote.
The composition of each clause is controlled by a vector of \emph{exclude} bits
(see \cref{fig:tm-overview}). These bits are parameters that are learned
by teams of \acp{TA}.

Each clause can produce a vote for its class. The algorithm states that half of the clauses vote
positively, while the other half of the clauses vote negatively (we will
denote this by the boolean \cneg{}). The
inclusion of inhibition in the voting system enables non-linear decision boundaries in the
inference process. A majority vote gives an indication of class confidence. This
confidence is used to classify the input data and influence future decisions of
the automata through the feedback mechanism~\cite{granmo2018tsetlin}. In this
work we consider only the single class \ac{TM} for simplicity.

The \ac{TA} is a class of finite reinforcement
automaton~\cite{narendra1989learning}. It produces an
\emph{exclude} output for states below the midpoint, and \emph{include} for
states above the midpoint as illustrated in \cref{fig:ta-state}. The \ac{TA}
receives a penalty or reward from the feedback mechanism based on the current
state of the \ac{TM}. Continued rewards in the end states cause the \ac{TA} to
saturate. A penalty in one of the midstates ($n$ or $n+1$) causes the \ac{TA} to
transition across the decision boundary---inverting its output from exclude to
include, or vice versa. The feedback mechanism is described in more detail in
\cref{sec:fb}.

\begin{figure}
        \centering
        \tikzinput{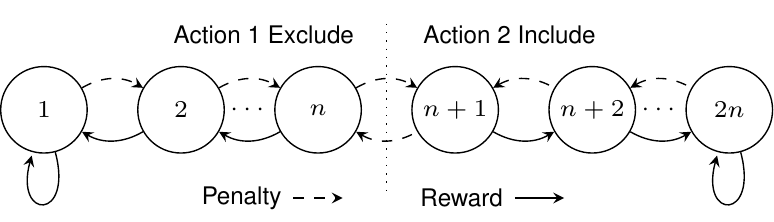}
        \caption{%
                \Acl{TA} state graph.
        }\label{fig:ta-state}
\end{figure}

	\section{Asynchronous Tsetlin Machine}\label{sec:tm}

\Cref{fig:tm-pn} is an architectural diagram for the asynchronous implementation of the
\ac{TM} shown in \cref{fig:tm-overview}, demonstrating the scalability of the design. The diagram can be
composed with the Petri-net-like tiles in \cref{fig:tm-cells} to form a
visualization of the complete system. In these diagrams, rectangles represent transitions or computations, and circles represent places as in classical Petri nets. When abutting these tiles,
adjoining places are reduced to a single place. This representation of the system is not a formalism.

\begin{figure}
        \centering
        \tikzinput{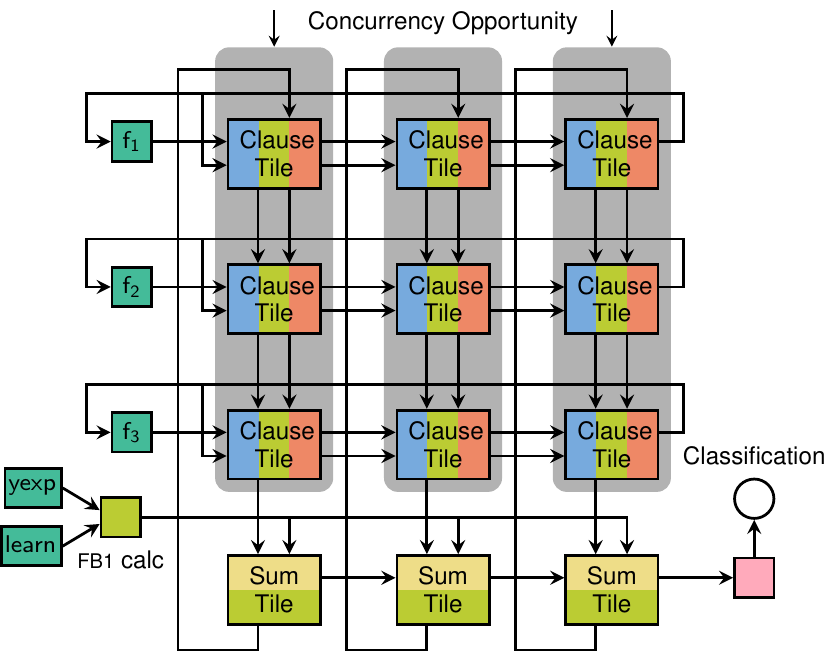}
        \caption{%
                Architectural diagram for a \acl{TM} with three \finputs{} (rows) and three
                clauses (columns).
        }\label{fig:tm-pn}
\end{figure}

\begin{figure}
        \centering
        \subfloat[Clause tile.]{\includegraphics[scale=\petriscale]{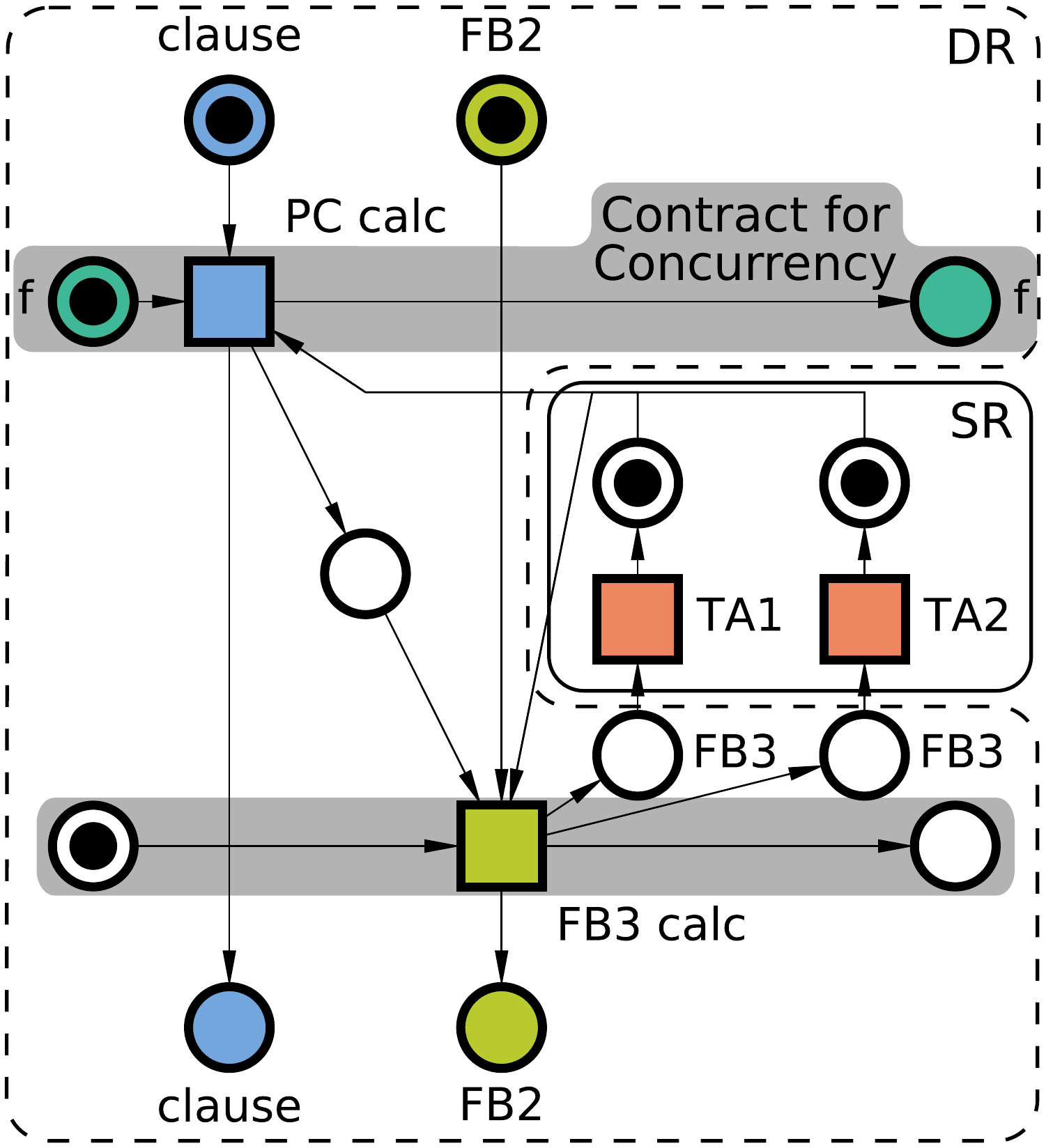}%
                \label{fig:pc-cell}}%
        \hfil
        \subfloat[Sum tile.]{\includegraphics[scale=\petriscale]{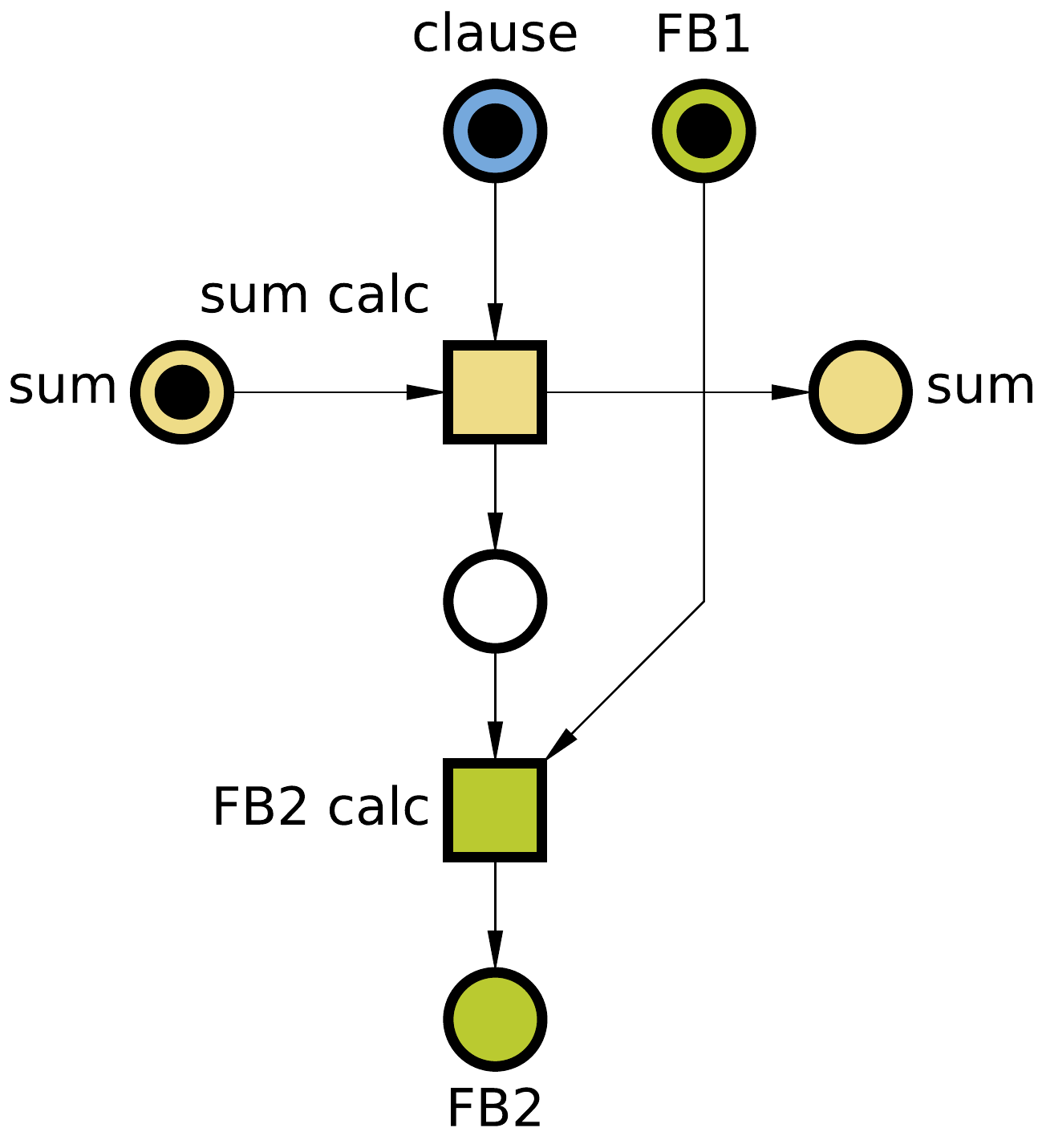}%
                \label{fig:sum-cell}}%
        \caption{%
                Petri-net-like visualizations for the \acl{TM} system:
                \protect\subref{fig:pc-cell} the clause tile including partial
                clause ({\smaller PC}) and two \acfp{TA};
                and \protect\subref{fig:sum-cell} the sum tile.
        }\label{fig:tm-cells}
\end{figure}

The main structure is a $3\times 3$ grid constructed with \finputs{} in rows and
clauses in columns. The small size facilitates the explanation of the system,
but the architecture can be easily extended by appending clause columns to the
right side, or feature rows below \signal[3]{f}.

The clause tile is detailed in \cref{fig:pc-cell} and comprises a partial clause
({\smaller PC}) computation for the \finput{} and its complement (as introduced
in \cref{sec:tm-intro}); a partial feedback calculation for stage three feedback
(\fbstage{3}); and two \acp{TA}. The feedback stages will be discussed later in
\cref{sec:rand}.

The clause sum is computed in the sum tile at the bottom of each clause column.
\Cref{fig:sum-cell} details the sum tile. Stage two feedback (\fbstage{2}) for
the clause is computed at this stage using the clause sum.

The diagrams capture the potential for both concurrency and serialization in the
design. As presented, the system operates in a serialized manner where the
\finputs{} are passed through each clause in series. By contracting the places
in the highlighted regions of \cref{fig:pc-cell}, tokens are passed to the right
without waiting for {\sffamily\textsmaller{PC} calc} and {\sffamily\fbstage{3}
calc}, this allowing clauses to compute in parallel.

The majority of the system is implemented in a \acf{DR} style using \iac{RCD}
scheme~\cite{wheeldon2019selftimed}. However, the \acp{TA} are implemented in a
\acf{SR} style (as indicated in \cref{fig:pc-cell}) which will be motivated and
discussed in \cref{sec:ta}.

The \ac{RCD} scheme is a simplified scheme which completion detects only the
outputs of a block. This brings benefits of early propagation and drastically
reduced area and power overheads versus traditional full completion detection.
The trade-off is an added timing assumption on the return-to-idle phase of the
\ac{DR} signals. Full details of the \ac{RCD} scheme are available
in~\cite{wheeldon2019selftimed}.

\subsection{Inference Circuits}

The inference datapath comprises the clauses, majority voting and classification
introduced in \cref{fig:tm-overview}. Clauses are separated into partial clause
calculations as discussed earlier in this section and can be implemented by the
schematics in \cref{fig:pc-schem,fig:comb-schem}. \Finputs{} occur in \ac{DR}
encoding along with \emph{exclude} signals \signal[0]{e} and \signal[1]{e}.
The outputs of the partial clause are
combined using \iac{DR} \gate*{and} tree.

Majority voting calculated the sum of the clause votes (\csum{}) and is
implemented using \ac{DR} \popcount{}. The schematic in
\cref{fig:popcount-schem} shows the implementation based on \ac{DR} half- and
full-adders. The \gates{or} and wires in this circuit are also implicitly
\ac{DR} encoded. Two spacer inverters (spinv) are required to ensure uniform
spacer polarity at the outputs.

For this single class \ac{TM} example, the output is classified using a
threshold function in the form of a magnitude comparator. The magnitude
comparator lends itself to low energy implementation in \ac{DR} and saves energy
by evaluating single-bit comparisons from \acl{MSB} to \acl{LSB} only if
needed, all within the period of one cycle~\cite{wheeldon2021lowlatency}. To expand the example to two classes or
more, an argmax function would take the place of thresholding.

\begin{figure}
        \centering
        \pbox{\linewidth}{%
                \relax\ifvmode\centering\fi
                \setcounter{subfigure}{0}%
                \subfloat[Partial clause (\textsmaller{PC}).]{\tikzinput{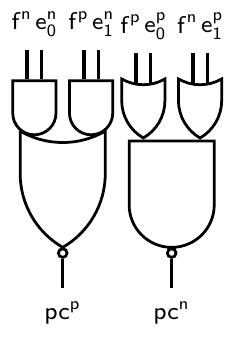}%
                        \label{fig:pc-schem}}\\%
                \subfloat[Partial clause combiner.]{\makebox[1.15\width][c]{\tikzinput{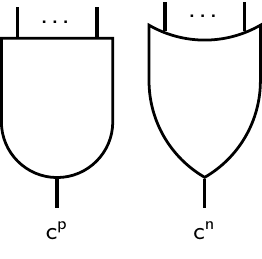}}%
                        \label{fig:comb-schem}}%
        }%
        \hfil
        \pbox{\linewidth}{%
                \setcounter{subfigure}{2}%
                \subfloat[\Popcount{} with implicit \acl{DR} wires and blocks.]{\tikzinput{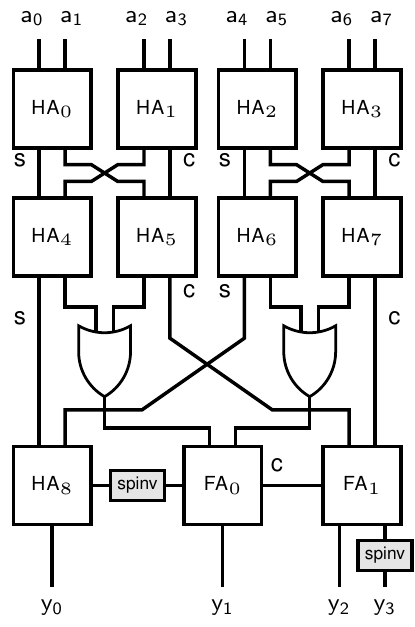}%
                        \label{fig:popcount-schem}}%
        }\\%
        \subfloat[Magnitude comparator.]{\includegraphics[scale=0.27]{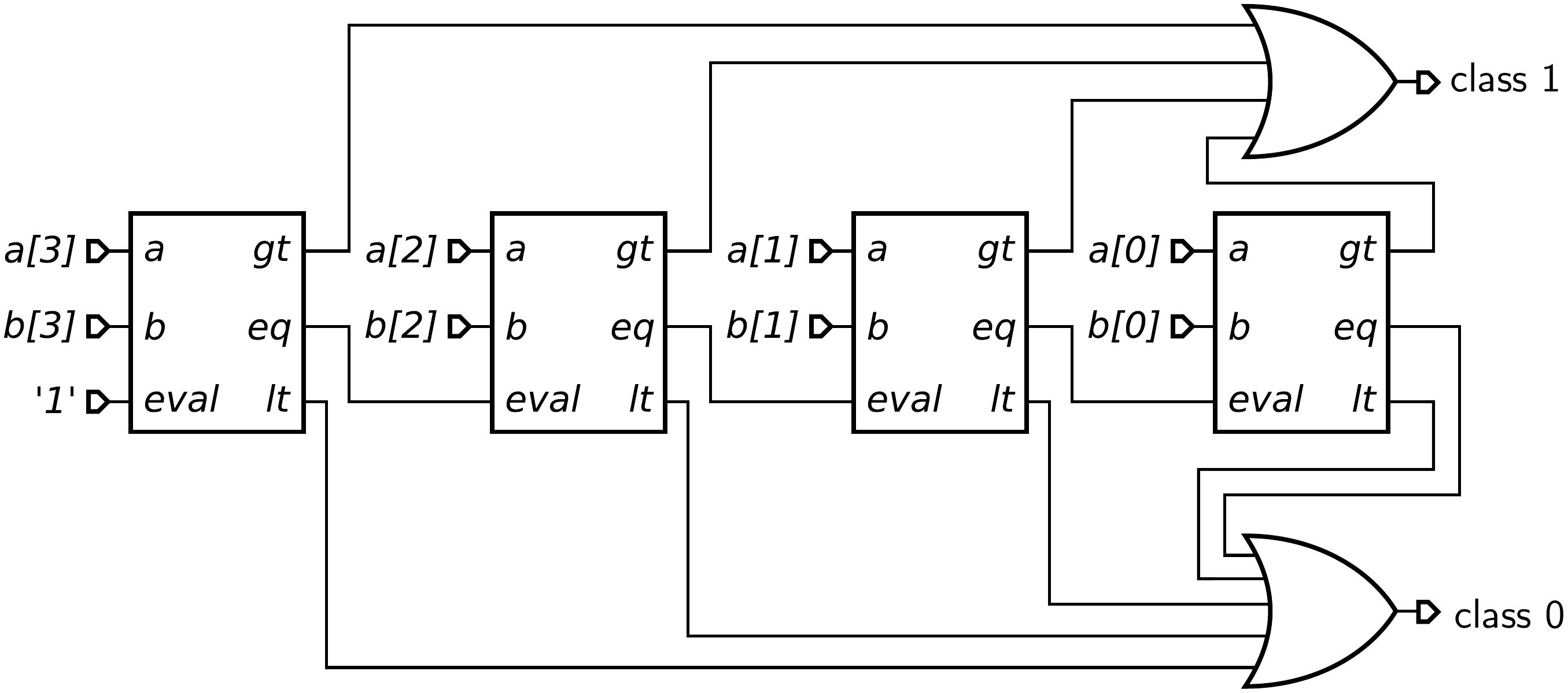}%
                \label{fig:comp-schem}}
        \caption{%
                \Acl{DR} circuits for inference.
                Spinv = spacer inverter.
        }\label{fig:infer-schem}
\end{figure}

In the following sections we will explore the inference circuit delays and
analyze the inference latency of the asynchronous datapath with differing input
operands.
For synthesis we use a commercial, off-the-shelf technology library: \processlibname{umc65ll},
herein referred to as \processlibnameshort{umc65ll}.

\subsection{Delay Analysis}

Since we use \iac{RCD} scheme which enables early
propagation~\cite{wheeldon2021lowlatency}, the delays of the
circuits can vary based on the input operands. We analyze the delay
distributions of the inference circuits using uniformly-distributed random
inputs to gain an understanding of this phenomenon. We will use this to explain
the evolution of latency during training in the next section.

\paragraph*{Clause} In the worst-case delay no literals are excluded
($\signal{e}=0$), all \finputs{} are logic-1 ($\signal{f}=1$); therefore the
delay is bounded by the slowest input and the delay arc is through the positive
rail. In the best-case delay, all but one literal is excluded and the literal is logic-0;
therefore the delay is bounded by signal propagation through the \gate*{or} tree
in the negative rail. Since the clause is a large \gate*{and} tree, there is a
high probability of $\signal{c}=0$, so the delay distribution is mostly
clustered at the lower end as \cref{fig:group-dist} shows. A much smaller
portion is seen towards the circuit's maximum delay and is associated with
$\signal{c}=1$.

\paragraph*{\PopCount{}} The delay of the circuit is bounded by the carry chain
formed by \bname[8]{ha}, \bname[0]{fa} and \bname[1]{fa}. In the best case there
are no carries and the circuit latency is governed by the longest logic path
which comprises the first two layers of \bname{ha}\textsmaller{s} (which have
roughly equal paths), through the \gates{or} and finally \bname[0]{fa} to the
\signal[1]{y} output. In the worst case, output \signal[3]{y} must wait for the
aforementioned best-case path as well as the carry through \bname[1]{fa}. This
leads to the log-normal-like delay distribution seen in \cref{fig:group-dist}.

\paragraph*{Magnitude Comparator} For uniform random inputs and a normalized
worst-case delay of $1$, the comparator achieves a mean delay of $0.05$. This
is attributed to early propagation and the evaluation of single-bit comparisons
from \acs{MSB} to \acs{LSB}. The delay of the circuit increases as the absolute
difference between operands decreases, because more of the \acsp{MSB} are equal.
This circuit is therefore fastest during inference and the later stages of
training. The operand space halves for each single-bit comparison leading to the
negative exponential delay distribution in \cref{fig:group-dist}.

\begin{figure}
        \centering
        \tikzinput{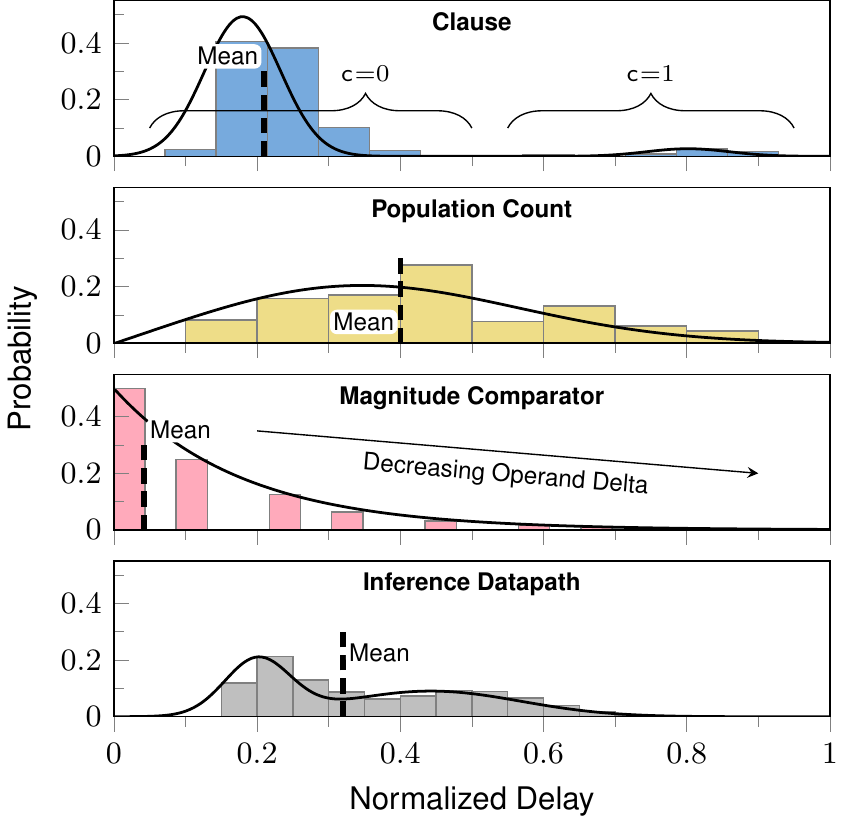}
        \caption{%
                Latency distributions of the \acl{TM} inference datapath and its
                components for uniformly-distributed random inputs.
                Post-synthesis results using the \processlibname{umc65ll} library.
        }\label{fig:group-dist}
\end{figure}

\subsection{Operand-dependent Inference Latency}

We investigate the evolution of inference latency during training by plotting
the \sptocw{} latency at different stages. We use a binarized version of the \pnoun{Iris}
Flower dataset\footnote{Available: https://github.com/cair/TsetlinMachine}
to provide feature inputs (\signal{f}). Exclude inputs (\signal{e}) are
retrieved from \iac{TM} pre-trained on the same dataset after the specified
number of epochs. \Cref{fig:infer-dist} shows the distributions for the
untrained \ac{TM}, and the trained \ac{TM} after 4 epochs and 50 epochs. The
plots show the trend of the mean towards greater circuit delay as the
\ac{TM} is trained and therefore more clauses become activated during the
computation of each datapoint ($\signal{c}=1$). In this case we see more
examples from the right side of the clause delay distribution in
\cref{fig:group-dist}. This is the main contributor to the increase in mean latency as the system becomes more trained. The magnitude comparator offers a net \emph{decrease} in mean latency as training progresses since the difference between input operands will increase, meaning we shift towards the left of the distribution in \cref{fig:infer-dist}. However this shift is somewhat smaller than that of the clauses, and it therefore has a lesser effect on the inference datapath. The distribution of the population count is much more uniform than that of the other circuits, and therefore has a negligible effect during training.

\begin{figure}
        \centering
        \tikzinput{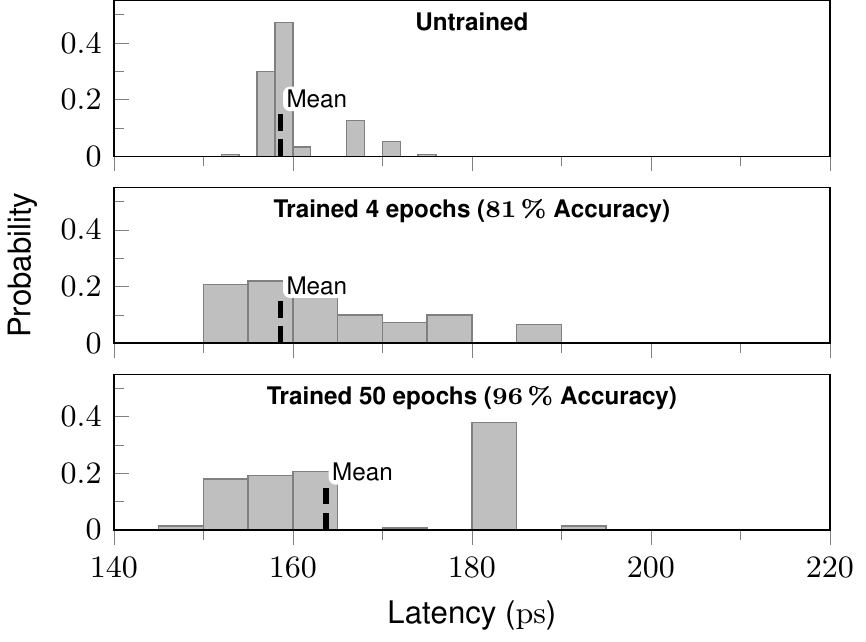}
        \caption{%
                Latency distribution of the \acl{TM} inference datapath for the
                \pnoun{Iris} dataset.
        }\label{fig:infer-dist}
\end{figure}

	\section{Decomposition of Feedback}\label{sec:fb}

During learning, feedback is given to each \ac{TA} to influence its future action, and therefore
control pattern detection in the \ac{TM} as a whole. The feedback is based on a
carefully-designed payoff matrix~\cite[p.~10]{granmo2018tsetlin} and is split
into two categories: Type I and Type II. Type I feedback reinforces good
decisions and penalizes false negative outputs. Type II feedback combats false
positives by penalizing \acp{TA} that exclude when $\signal{c}=1$.

We decompose the feedback logic into three stages based on scope:
\fbstage{1} at \ac{TM} level,
\fbstage{2} at clause level,
and \fbstage{3} at \ac{TA} level.
\fbstage{1} and \fbstage{2} are assigned types: \fbtype{0}, \fbtype{1} or
\fbtype{2}; these are based on the Type I and Type II concepts introduced
previously.
\fbstage{3} assigns an action to a single \ac{TA}: \taaction{1}, \taaction{2} or
\taaction{0}; and is based on the outcomes of the payoff matrix%
\footnote{We do not consider boosting of true positive feedback, which is an
option in the original algorithm.}.
Each feedback stage's output is input to the next, until finally the output of \fbstage{3}
is input to the corresponding \ac{TA}.
The following sections discuss each feedback stage in detail.
Finally we will discuss the synthesis of the feedback circuits in the \ac{DR}
style in \cref{sec:fb-impl}.

\subsection{\fbstage{1}: \acs{TM}-level Feedback}\label{sec:fb-tm}
    \emph{Inputs:} \signal{learn}, \yexp{}\\
    \indent\emph{Output:} Feedback type (\fbtype{0}, \fbtype{1} or \fbtype{2})

    \noindent
    The first stage broadly enables feedback for the entire \ac{TM} if
    \signal{learn} is asserted. The feedback type is chosen according to
    \signal{yexp}---the expected class as provided by the labeled data---as follows:
    \[
        \fbstage{1} = 
        \begin{cases}
            \fbtype{0}, & \text{for }\signal{learn} = 0\\
            \fbtype{1}, & \text{for }\signal{learn} = 1, \yexp{} = 1\\
            \fbtype{2}, & \text{for }\signal{learn} = 1, \yexp{} = 0\\
        \end{cases}
    \]

\subsection{\fbstage{2}: Clause-level Feedback}\label{sec:fb-clause}
\emph{Inputs:} \fbstage{1}, \cneg{}, \fbprobsig{2}\\
\indent\emph{Output:} Feedback type (\fbtype{0}, \fbtype{1} or \fbtype{2})

\noindent
The second stage swaps the feedback type from \fbtype{1} to \fbtype{2}, or vice
versa if the associated clause is negated ($\cneg = 1$, discussed in
\cref{sec:tm-intro}).
Additionally, feedback is stopped (inaction given) if the number of activated
clauses (those producing a logic-1 at the output) meets the threshold,
\tparam{}. This is determined by probabilistic choice, \fbprobsig{2}, which is
discussed at the end of this section. If there is no feedback from the previous
stage ($\fbstage{1} = \fbtype{0}$), then none will be generated ($\fbstage{2} =
\fbtype{0}$). Parameter \cneg{} is determined at design time, therefore the
logic for \fbstage{1} can be separated and depends only on \fbstage{1} and
\fbprobsig{2}. This will be demonstrated in \cref{sec:fb-impl}.
Clause-level feedback can be summarized as follows:
\[
    \fbstage{2} = 
    \begin{cases}
        \fbtype{0}, & \text{for }\fbstage{1} = \fbtype{0}\\
        \fbtype{0}, & \text{for }\fbstage{1} = \fbtype{1}, \fbprobsig{2} = 0\\
        \fbtype{0}, & \text{for }\fbstage{1} = \fbtype{2}, \fbprobsig{2} = 1\\

        \fbtype{1}, & \text{for } \cneg{} = 1, \fbstage{1} = \fbtype{2}, \fbprobsig{2} = 0\\
        \fbtype{2}, & \text{for } \cneg{} = 1, \fbstage{1} = \fbtype{1}, \fbprobsig{2} = 1\\

        \fbtype{1}, & \text{for } \cneg{} = 0, \fbstage{1} = \fbtype{1}, \fbprobsig{2} = 1\\
        \fbtype{2}, & \text{for } \cneg{} = 0, \fbstage{1} = \fbtype{2}, \fbprobsig{2} = 0\\
    \end{cases}
\]

The \ac{TM} algorithm describes probabilities of generating \fbtype{1} and
\fbtype{2}~\cite[p.~10]{granmo2018tsetlin}.
In this work we refer to them as \fbprob{1} and \fbprob{2} which are based on
\csum{} and \tparam{} as follows:
\[
    \fbprob{1} = \frac{\tparam - \text{clamp}(\csum)}{2\tparam},\ 
    \fbprob{2} = \frac{\tparam + \text{clamp}(\csum)}{2\tparam},
\]
where the clamp function restricts its argument to $[-\tparam,\tparam]$
such that the probabilities lie in the range $[0,1]$.
As the number of activated clauses (those producing $c=1$) approaches \tparam{}, the probability of
the \acp{TA} in that clause receiving feedback decreases.
In order to generate a randomized choice based on these probabilities, we could assign
booleans as follows:
\[
    \signal[1]{q} = \text{rand} < \fbprob{1},\ 
    \signal[2]{q} = \text{rand} < \fbprob{2}
\]
where rand is a real number in the range [0,1].
Since \fbprob{1} and \fbprob{2} are complementary in the range
[0,1], \signal[1]{q} and \signal[2]{q} are also complementary in the binary
domain and therefore $\signal[2]{q} = \bnot{\signal[1]{q}}$.
We therefore introduce a single new signal for \fbstage{2}, \fbprobsig{2},
which takes the place of the preceding probabilities. The generation of this
signal will be discussed in \cref{sec:p2gen}.

\subsection{\fbstage{3}: \acs{TA}-level Feedback}\label{sec:fb-ta}
\emph{Inputs:} \fbstage{2}, \signal{inc}, \signal{c}, \signal{x}, \fbprobsig{3}\\
\indent\emph{Output:} \ac{TA} action (\taaction{0}, \taaction{1} or \taaction{2})

\noindent
The third and final feedback stage translates the feedback type from
\fbstage{2} into a \ac{TA} action for the specified \ac{TA} based on:
the \ac{TA}'s current action, encoded as \emph{include} in signal
\signal{inc};
the current clause output, \signal{c};
the feature of complemented feature (\signal{f} or \nsignal{f}) associated with the \ac{TA}, \signal{x};
and a probabilistic choice component, \fbprobsig{3}.

The \fbstage{3} output is chosen according to \cref{tab:fb3},
where $\fbprobsig{3} = 1$ denotes that the higher probability
$(\sparam-1) / \sparam$ option is chosen from the payoff matrix.
$\fbprobsig{3} = 0$ denotes that the lower probability
$1 / \sparam$ is chosen. The generation of signal \fbprobsig{3} will be discussed
in \cref{sec:p3gen}.

\begin{table}
    \setlength{\tabcolsep}{0.1em}
    \newcolumntype{f}{>{\collectcell\fbtype}l<{\endcollectcell}}
    \newcolumntype{t}{>{\collectcell\taaction}l<{\endcollectcell}}
    \newcolumntype{E}[1]{>{\collectcell\fillempty}#1<{\endcollectcell}}
    \newcommand{\fillempty}[1]{\if\relax\detokenize{#1}\relax\ensuremath{\times}\else#1\fi}
    \centering
    \caption{%
        Truth table for \fbstage{3}. $\times=\text{don't care}$.
    }\label{tab:fb3}
    \begin{minipage}[t]{.5\linewidth}
        \centering
        \begin{tabular}[t]{f @{\hskip.5em} *{4}{E{C}} @{\hskip1em} | @{\hskip1em} t}
            \toprule
            \multicolumn{1}{l}{\fbstage{2}} &
            \multicolumn{1}{c}{\signal{inc}} &
            \multicolumn{1}{c}{\signal{c}} &
            \multicolumn{1}{c}{\signal{x}} &
            \multicolumn{1}{l}{\fbprobsig{3}} &
            \multicolumn{1}{l}{\fbstage{3}} \\\midrule
            0 & & & & & 0 \\[1ex]
            1 & 1 & 0 &   & 0 & 1 \\
            1 & 1 & 0 &   & 1 & 0 \\
            1 & 1 & 1 &   & 0 & 0 \\
            1 & 1 & 1 &   & 1 & 2 \\
            1 & 0 & 0 &   & 0 & 2 \\
            1 & 0 & 0 &   & 1 & 0 \\
            \bottomrule
        \end{tabular}
    \end{minipage}%
    \begin{minipage}[t]{.5\linewidth}
        \centering
        \begin{tabular}[t]{f @{\hskip.5em} *{4}{E{C}} @{\hskip1em} | @{\hskip1em} t}
            \toprule
            \multicolumn{1}{l}{\fbstage{2}} &
            \multicolumn{1}{c}{\signal{inc}} &
            \multicolumn{1}{c}{\signal{c}} &
            \multicolumn{1}{c}{\signal{x}} &
            \multicolumn{1}{l}{\fbprobsig{3}} &
            \multicolumn{1}{l}{\fbstage{3}} \\\midrule
            1 & 0 & 1 & 0 & 0 & 0 \\
            1 & 0 & 1 & 0 & 1 & 2 \\
            1 & 0 & 1 & 1 & 0 & 0 \\
            1 & 0 & 1 & 1 & 1 & 1 \\[1ex]
            2 & 1 &   &   &   & 0 \\
            2 & 0 & 1 & 0 &   & 1 \\
            2 & 0 & 0 &   &   & 0 \\
            \bottomrule
        \end{tabular}
    \end{minipage}
\end{table}

\subsection{Synthesis of Feedback Circuits}\label{sec:fb-impl}

Using one-hot encoding in the feedback circuits maintains
direct compatibility with the inference datapath and
ensures speed independence, which is important for hazard-freeness.
Signals \signal{inc}, \signal{c}, \signal{x} are already one-hot encoded (\dr{})
for the inference circuits in \cref{sec:tm}, so these encodings are reused
here in the learning circuits.
Furthermore, we encode \fbstage{1} as follows, and \fbstage{2} similarly:
$\fbstage{1}=\{ \fbrail{1}{2}, \fbrail{1}{1}, \fbrail{1}{0} \}$.
\fbstage{3} describes \ac{TA} actions and is encoded as:
$\fbstage{3}=\{ \fbraila{3}{2}, \fbraila{3}{1}, \fbraila{3}{0} \}$,
for \taaction{2}, \taaction{1}, \taaction{0} respectively. Note that rail
orders do not matter as we refer to the rails by name.
In this way, to indicate \fbtype{2} feedback on \fbstage{1}, we would set
$\fbrail{1}{2}=1, \fbrail{1}{1}=0, \fbrail{1}{0}=0$.

Using the previously defined logic, we can generate circuits for the feedback.
The implementation of stage two
feedback depends on the clause polarity---it being either negated ($\cneg=1$) or
non-negated ($\cneg=0$). Note that the stage one circuit is instantiated \emph{once
per \ac{TM}}, the stage two circuit \emph{once per clause}, and the stage three circuit
\emph{once per \ac{TA}}.

In stage three, the \fbstage{2} signal is shared throughout the \acp{TA} within
the same clause. For this reason, we carefully design the logic so that
\fbstage{2} signals are injected as close to the outputs as possible, making the
propagation path the shortest. Therefore when $\fbstage{2}=0$,
the computation on the corresponding \fbstage{3} rails will conclude rapidly,
and for all \acp{TA} in the clause.

	\section{Generation of Random Bits \fbprobsig{2}, \fbprobsig{3}}\label{sec:p2gen}\label{sec:p3gen}\label{sec:rand}

We introduced two distinct probabilistic choice mechanisms in \cref{sec:fb}.
In \cref{sec:fb-clause} we introduced the signal \fbprobsig{2}: a probabilistic boolean
chosen at the clause level which may force $\fbstage{2}=\fbtype{0}$. This boolean is
dependent on \tparam{} and \csum{}. The probability of $\fbprobsig{2}=1$ varies
at runtime according to \csum{}.
In \cref{sec:fb-ta} we introduced the signal \fbprobsig{3}, which is a random
boolean generated for each \ac{TA}. It is required under certain circumstances to
choose between two \ac{TA} actions. This boolean is dependent on \sparam{} and is
fixed during runtime.

From these requirements, we must be able to generate
\emph{biased} random bits: bits where $\prob{1} \not= 0.5$ (\emph{unbiased}
random bits have $\prob{1}=0.5$, such as those generated by \iac{LFSR}). In
addition, we must be able to vary \prob{1} at runtime for \fbprobsig{2}.
Probabilistic choices are used in the \ac{TM} to ensure diversity of learning.
A pseudorandom generator is sufficient to satisfy these needs, and benefits from higher energy
efficiency than a \emph{true} random generator.

Due to the requirement of \acp{PRBG} for each \ac{TA}, and additionally for each
clause, their area and energy consumption are of utmost importance.
We chose to implement \iac{PRBG} based on the principles of irregular sampling of a
regular waveform, as this allows us to minimize the overheads and take advantage
of the asynchronous nature of the rest of the system.

\subsection{Asynchronous Sampling of Clock with Variable Duty Cycle}

We take advantage of the asynchronous inference logic by using
asynchronous handshakes to sample a regular clock
waveform. A clock with \SI{50}{\percent} duty cycle will generate unbiased
random bits. A clock can be generated using \iac{RO}. To ensure random sampling,
the clock and handshake must be \emph{uncorrelated}. \Acp{RO} can be gated by
adding a \gate*{nand} or \gate{nor} into the ring. However this technique will
disadvantage us in this case as the \ac{RO} will always start up in the same
phase, therefore the clock and sampling signals may become correlated.

We take advantage of the properties of a purely inverter-based \ac{RO}. The
\ac{RO} can be power gated using header\slash footer transistor when entropy
generation is not required. Such \iac{RO} will start up in a non-deterministic
phase according to thermal noise and other effects in the inverters.

To generate biased bits for \fbprobsig{2} and \fbprobsig{3}, we need to vary the
duty cycle of the clock. \citet{agustin2015indepth} show how to
construct \iac{RO} such that each tap has a unique duty cycle. This is
achieved by using inverters with asymmetric rise\slash fall times. For example,
odd inverters have fast rise and slow fall times, and even inverters vice versa.
This can be done via transistor sizing (\ie{} in the silicon library), or
altering supply voltage to the inverters (\ie{} at implementation stage).
Using this \ac{RO} setup, we can multiplex between taps to obtain different
clock duty cycles, and therefore alter our \ac{PRBG} probability at runtime.

We have introduced a clock into our asynchronous circuit, and this may seem
counterintuitive, however the load on this clock (and therefore its energy
consumption) will be low compared with fully synchronous designs where the clock
drives large numbers of flip-flops. In \cref{sec:rand-no} we will investigate
the optimal number of \acp{PRBG} required. By taking this distributed \ac{PRBG}
approach, we can vastly reduce the circuit area compared with a na\"ive approach
using one \ac{LFSR} per \ac{TA}.

\Cref{fig:handshake} shows the circuit used to select the correct duty cycle
clock and sample it with the asynchronous \signal{req} input. The mutex ensures
\prail{ack} and \nrail{ack} outputs are mutually exclusive. These outputs
represent the random output bit with a \dr{} encoding. That is for
$\{\prail{ack},\nrail{ack}\}$: $\{0,1\}$ represents logic-0, and
$\{1,0\}$ logic-1. $\{0,0\}$ is the \emph{spacer} or \emph{null} state
used to separate valid output values temporally.

When the \signal{clk} and \signal{req} signals are both low, the output will be
in the spacer state. If a request is made (by asserting \signal{req}) during the low
period of the clock, the output of the set-dominant latch is reset, and
\signal{req} wins the mutex, resulting in \nrail{ack} asserting. If
\signal{clk} rises now, the output of the latch rises, however the mutex is
still held by \signal{req}. A following deassertion of \signal{req} will result
in \signal{clk} gaining the mutex, however \prail{ack} will be masked by the
\gate{and}.

If a request is made during the high period of \signal{clk}, the output of the
latch will already be high and \signal{clk} will have won the mutex. On the rise
of \signal{req} the \gate{and} will unmask \prail{ack}. If \signal{clk} falls
now, the output of the latch remains high since the \textsf{\textsc{r}} input is inactive, and
therefore \prail{ack} also remains high.

The circuit is synthesized for the \processlibnameshort{umc65ll} cell library
and the results are summarized in \cref{tab:results-rand}. The asynchronous
handshake shows almost $10\times$ saving in area compared to an 8-bit \ac{LFSR}. Power
and energy are also drastically decreased. The
asynchronous handshake is well suited to energy-conscious, pervasive
applications.

\begin{figure}
    \centering
    \tikzinput{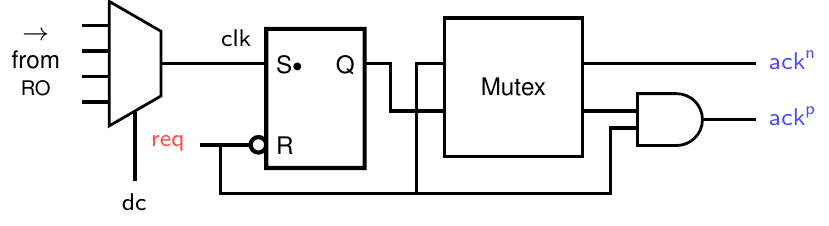}
    \caption{%
        Handshake circuit for the asynchronous \acl{PRBG} with \acl{DR} output.
    }\label{fig:handshake}
\end{figure}

\subsection{Optimal Number of \acsp{PRBG}}\label{sec:rand-no}

The original \ac{TM} algorithm takes a probabilistic choice for every \ac{TA}
update involving \sparam{}. However \citet{abeyrathna2020novel} show that
randomizing every \nth{d} \ac{TA} update can save energy with a minimal drop in
test accuracy. With $d = 1000$, an accuracy within \SI{2}{\percent} of $d=1$ was
maintained for three of the four datasets tested.

Following on, we could hypothesize an optimal number of \acp{PRBG} to fulfill
the needs of a given \ac{TM}. If every \nth{100} \ac{TA} update is randomized,
we will need to produce at most $\Ntas / 100$ probabilistic choices per learning
phase.

This could result in a $100\times$ reduction in \acp{PRBG} for
\iac{LFSR}-based design, in exchange for arbitration overhead to share one
\ac{PRBG} between multiple \acp{TA}. Similar could be said for the asynchronous
handshake design, although the already small size of the design means that
arbitration overhead would almost certainly result in a larger area than
instantiating an asynchronous handshake \ac{PRBG} for every \ac{TA}.

\begin{table}
        \centering
        \caption{%
                \Acp{PRBG} synthesized for \processlibname{umc65ll}
                silicon library.
        }\label{tab:results-rand}
        \begin{spreadtab}{{tabular}{%
            @{}
            l
            S[table-format=2.2]
            S[table-format=1.2]
            S[table-format=1.3]
            S[table-format=1.3, round-mode=figures, round-precision=3]
            @{}
        }}
                \toprule
                @ Implementation &
                    @ {Area} &
                    @ {Cycle Time [\si{\nano\second}]} &
                    @ {Power [\si{\micro\watt}]} &
                    @ {Energy [\si{\femto\joule}]} \\
                \midrule
                @ LFSR8     & 92.2 & 0.38 & 6.41  & c2*d2 \\
                @ Handshake & 9.72 & 0.23 & 0.621 & c3*d3 \\
                \bottomrule
        \end{spreadtab}
\end{table}

	\section{Asynchronous Tsetlin Automaton}\label{sec:ta}

In a \ac{TM}, the \acp{TA} consume most of the hardware
resources~\cite{wheeldon2020learning}. The number of \acp{TA} is given by
$\Ntas=2 \Nclasses \Nclauses \Ninputs$. Therefore the goal is to create a
lightweight \ac{TA} implementation with a focus on area and energy. The \ac{TA}
hardware must implement the finite state automaton described by the state
diagram in \cref{fig:ta-state}.

This section explores two methods for the design of \iac{TA}. Firstly \iac{QDI}
approach starting from \iac{STG} specification; and secondly \iac{BD}
approach starting from a traditional synchronous specification, using matched
delays to time the design. The designs will be compared with a synchronous
implementation.

\subsection{Quasi Delay Insensitive Approach}\label{sec:ta-si}

The \ac{STG} in \cref{fig:ta-stg} represents a one-hot \ac{FSM} for a two
action, six-state \ac{TA} (for the definition of \acp{STG},
see~\cite{cortadella2002logic}). The inputs \signal{p} and \signal{r} come from
the \fbstage{3} rails \fbraila{3}{1} and \fbraila{3}{2} which were introduced in
\cref{sec:fb-impl}. The \fbraila{3}{0} rail denotes inaction and is only used
for completion detection.

State bits are encoded in
the \signal[**\_0]{x} and \signal[**\_1]{x} places at the top of the graph.
States \signal[1*]{x} (\signal[2*]{x}) are the action 1 (action 2) states---in
these states the automaton is indicating the action to exclude (include) the feature or feature complement from
the clause composition. The \signal[*1]{x} (\signal[*3]{x}) states are the
closest to (furthest from) the action decision boundary. As an example, when
\signal[13\_1]{x} holds a token, the automaton is indicating action 1 (exclude)
strongly.

A token at \signal[0]{p} denotes the idle state for the automaton.
It offers a choice between penalty and reward represented by signal
transitions \transplus{p} and \transplus{r} respectively.
Each subsequent branch from \transplus{p}, \transplus{r}, enables a
transition in one of the action output signals \signal[1]{a},
\signal[2]{a}.
Depending on the current action and the previous state, a state
transition may also occur.
The correct branch is chosen based on read arcs from the \signal[**\_1]{x}
places.
For example in the \nth[nd]{2} left-most branch of \transplus{r}, we transition from
state \signal[12]{x} to \signal[13]{x}.
This happens by firstly giving the token from place \signal[13\_0]{x} to
\signal[13\_1]{x}, and secondly giving token \signal[12\_1]{x} to
\signal[12\_0]{x}.
After state transitions and action output have occurred, the acknowledge
output, \signal{ack}, makes a positive transition. Finally, the previously
given input and action signals return low, followed by \transminus{ack}
and the token returns to \signal[0]{p}. The automaton is now ready for the next
penalty or reward input.

The internal \emph{transitions} \signal[**]{x} encode some information about the
states in the \ac{STG}. However, this is not enough to satisfy \ac{CSC} required
for synthesis of the \ac{STG} to logic gates~\cite{cortadella2002logic}. In the
previous example where we transition from \signal[12]{x} to \signal[13]{x},
there is a point where tokens are held by both \signal[13\_1]{x} and
\signal[12\_1]{x}. The \ac{STG} reaches this same state when transitioning
in the opposite direction, from \signal[12]{x} to \signal[13]{x}.

In order to achieve \ac{CSC}, we introduce internal signals to encode the
direction of travel of the state: \signal[*L*]{x}, \signal[*R*]{x}; for left and
right.
We insert these signals into every branch to maintain uniformity in the
\ac{STG}. Although not all these internal signals are strictly required, they
ease scalability of the \ac{STG} and help the synthesis tool to minimize and
share logic efficiently.

\begin{figure*}
        \centering
        \includegraphics[scale=\stgscale]{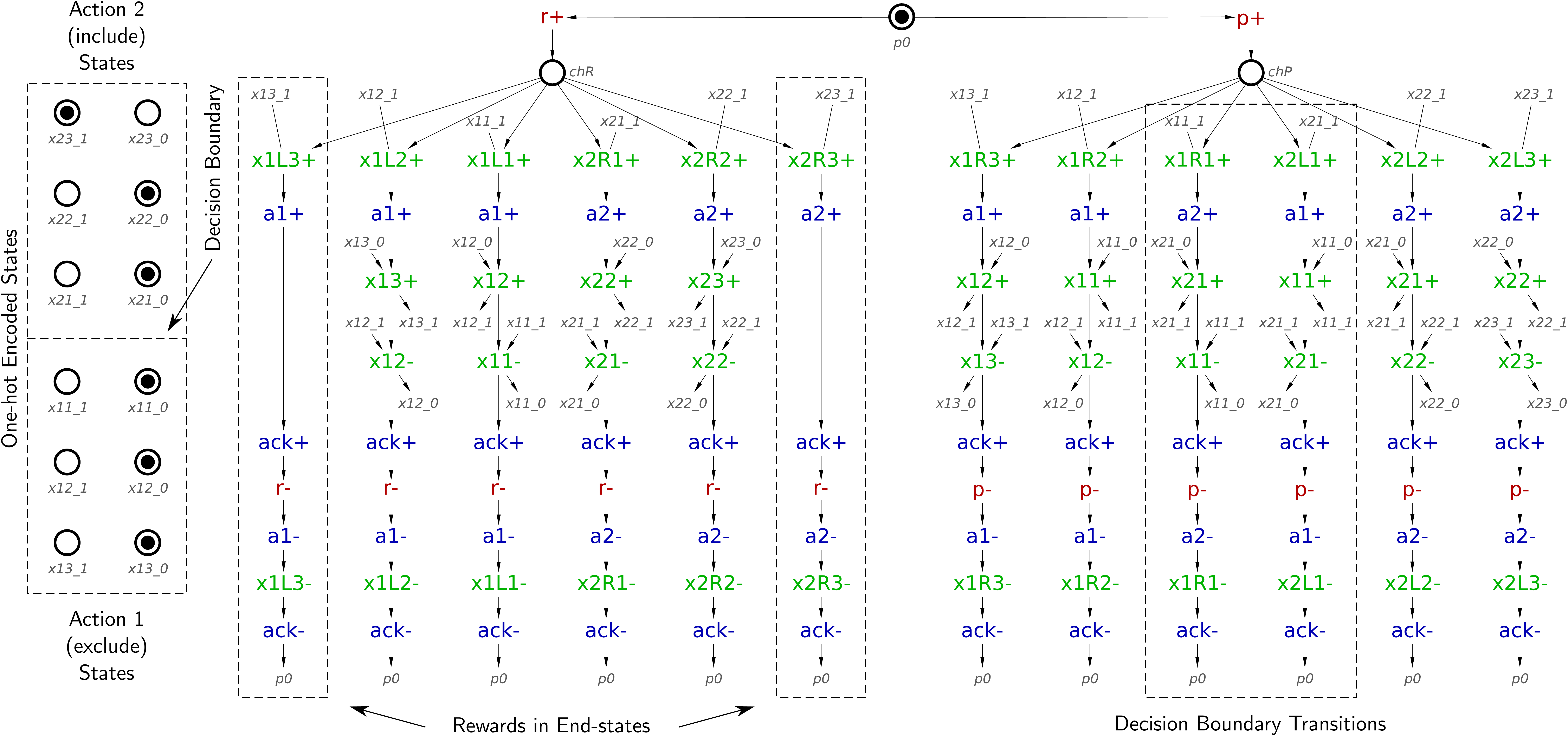}
        \caption{%
                \Acl{STG} for a \acl{TA} with two actions and six states.
                Inputs: \signal{p}, \signal{r}. Outputs: \signal[1]{a},
                \signal[2]{a}, \signal{ack}. Other signals are internal.
        }\label{fig:ta-stg}
\end{figure*}

The \ac{STG} in \cref{fig:ta-stg} passes all verification tasks required for
synthesis of a correct and hazard-free circuit in the \pnoun{Workcraft}
framework:
consistency, deadlock freeness, input properness, and output persistency.
The circuit was synthesized and mapped to the target technology by the
\pnoun{MPSat} backend (the \pnoun{Petrify} backend did not produce a result in a
reasonable amount of time).

In order to read the state of the \ac{TA} without giving penalty or reward
(during inference for example), we add a latch to the output to store one of the
\signal{p} or \signal{r} signals. The latch is controlled by the \signal{ack}
handshake.

The \ac{STG} state space expands exponentially as more signal transitions are
added. This makes it difficult for the synthesis tool to handle \acp{TA} with
more than three action states. Additionally there is a high possibility for
human error when copying and editing branches of the \ac{STG}. And as the number
of \ac{TA} action states increases, the graph becomes large and unwieldy to
navigate. For these reasons we explore a \ac{BD} approach to asynchronous
\ac{TA} design.

\subsection{Bundled-Data Approach}\label{sec:ta-md}

A synchronous one-hot \ac{FSM} can be constructed using flip-flops as storage
elements.
The next state logic can be derived using the adjacent states and \signal{p}, \signal{r} inputs as follows:
\begin{center}
\begin{minipage}{.5\linewidth}
        \noindent
        \begin{alignat*}{2}
                \signal[13]{x} &= \signal[13]{x} \band \signal{r} && \bor \signal[12]{x} \band \signal{r}\\
                \signal[12]{x} &= \signal[11]{x} \band \signal{r} && \bor \signal[13]{x} \band \signal{p}\\
                \signal[11]{x} &= \signal[12]{x} \band \signal{p} && \bor \signal[21]{x} \band \signal{p}
        \end{alignat*}%
\end{minipage}%
\begin{minipage}{.5\linewidth}
        \noindent
        \begin{alignat*}{2}
                \signal[21]{x} &= \signal[22]{x} \band \signal{p} && \bor \signal[11]{x} \band \signal{p}\\
                \signal[22]{x} &= \signal[21]{x} \band \signal{r} && \bor \signal[23]{x} \band \signal{p}\\
                \signal[23]{x} &= \signal[23]{x} \band \signal{r} && \bor \signal[22]{x} \band \signal{r}
        \end{alignat*}%
\end{minipage}
\end{center}

This synchronous circuit is converted into a self-timed one through the process
of desynchronization~\cite{cortadella2006desynchronization}, by decomposing the
flip-flops into master\slash slave latches and replacing the clock with
handshake controllers and delays matched to the combinational logic paths. The
timing of the resultant \ac{BD} design is not as robust as the \ac{QDI}
approach in \cref{sec:ta-si} since the matched delays are fixed at design time
and are subject to \ac{PVT} variations.

To reduce the overhead of the handshake controllers, we group the \acp{TA} into
an array and use one handshake controller for many latches.
Depending on the overall size of \ac{TM} for the target
application, the \acp{TA} could be grouped in different ways.
A monolithic array will minimize controller overhead, however there comes a
limit where buffers must be introduced into the latch enable signals to maintain
integrity. At this point, it may become desirable to group the \acp{TA} by clause.

The specification of the \ac{BD} design is easily scaled due to the use
of parameterizable Verilog code. The workflow takes advantage of mature
synchronous tools for synthesis.

\subsection{Synthesis Results}

\begin{table}[b]
        \centering
        \caption{%
                Results for the six-state \Acl{TA} synthesized for the
                \processlibname{umc65ll} silicon library.
        }\label{tab:results-ta}
        \begin{spreadtab}{{tabular}{%
                @{}
                l
                S[table-format=3.1]
                *{3}{S[table-format=1.2]}
                S[table-format=1.2, round-mode=figures, round-precision=3]
                @{}
        }}
                \toprule
                @ Implementation &
                        @ {Area} &
                        @ {\makecell{Cycle Time\\ {[\si{\nano\second}]}}} &
                        @ {\makecell{Leakage\\ {[\si{\nano\watt}]}}} &
                        @ {\makecell{Power\\ {[\si{\micro\watt}]}}} &
                        @ {\makecell{Energy\\ {[\si{\femto\joule}]}}} \\
                \midrule
                @ Synchronous  & 87.8 & 0.35 & 3.08 & 4.58 & c2*e2 \\
                @ Bundled-data\footnotemark & @86.0 & 0.41 & 2.89 & 5.19 & c3*e3 \\
                @ \Acs{QDI}    & 123  & 3.35 & 3.42 & 1.58 & c4*e4 \\
                \bottomrule
        \end{spreadtab}
        \\[0.5ex]
        \rlap{\hspace{1em}\footnotemark[4]Averaged over the array.}
\end{table}

Timing and power analyses of asynchronous circuits need special consideration.
In conventional synchronous \ac{STA}, delay through logic gates is calculated
using the input slew and output load capacitance.
Power estimations also rely on these parameters. Synchronous tools
cut combinational loops arbitrarily, therefore may be optimistic about slews,
gate delays and internal power in asynchronous circuits which contain such loops.

To obtain results for the \ac{QDI} design we use a specialized asynchronous
\ac{STA} tool~\cite{simoglou2020graphbased} which takes into account the
combinational loops in the timing calculation. The tool operates on the gate-level netlist and propagates slews
iteratively through circuits with loops, therefore avoiding optimism in the
slews and gate delays.
In the case there are multiple paths in the circuit which begin and end at the
same input\slash output combination, the tool uses the worst-case path to
determine the cycle time.

Since the \processlibnameshort{umc65ll} library does not
contain \cels{}, the \ac{QDI} circuit is combinational with feedback loops.
\Cref{tab:results-ta} shows a results summary of the \ac{TA} implementations.
All results are synthesis estimations and no layout is performed.
In this case, the \ac{QDI} circuit trades increased cycle time and energy for
increased robustness to \ac{PVT} variations. The \ac{BD} design strikes a good
trade-off between performance, area and elastic timing to integrate with the
rest of the system.
The layout process will introduce more power and latency into the synchronous
and \ac{BD} designs due to their clock trees, which the \ac{QDI} design will not
suffer from.

	\section{Conclusion}\label{sec:conclusion}

Our Petri net visualization of the \ac{TM} architecture enables flexible hardware
implementations by encapsulating both concurrency and serialization. By
decoupling the \ac{TA} storage elements from the clauses, \eg{} in a standalone
array, the \ac{TM} could be serialized further by reusing clause and sum
hardware across multiple cycles, trading off throughput for reduced logic area
(and leakage power).

Early propagation enabled by the \ac{RCD} scheme used in the \ac{DR} circuits
can be leveraged to trigger fine-grained power gating in the system to further
reduce energy consumption.

Distributed generation of probabilistic choices maximizes concurrency and decreases
energy overheads compared to a centralized approach. A simple \ac{PRBG} implementation
is sufficient and most suitable for our applications. Features of a clock source
undesirable in synchronous circuits (jitter, etc.) can be leveraged for
stochasticity.
Further study of \ac{PRBG} properties and their effects on \ac{TM} learning are
a subject for future work.

	\section*{Acknowledgment}

The authors thank Christos Sotiriou and colleagues for help in asynchronous \acs{STA} and
for access and support for their tool.
We are also grateful to colleague Thomas Bunnam for fruitful discussions on
\acsp{RO} and \acs{RNG} techniques.
The authors would like to acknowledge the funding support from EPSRC IAA grant: Whisperable.

	\bibliographystyle{IEEEtranN}
	\bibliography{ieee-config,zotero}
\end{document}